\documentstyle[namedreferences]{kluwer} 
\def\verbfont{\small\tt} 
\def \kkv{\vec{k}} 
\def \kr{cosmic ray} 
 
\runningtitle{KLUWER STYLE FILE} 
\runningauthor{G. MICHA{\L}EK ET AL.} 
 
\begin{opening} 
 
\title{Cosmic Ray Momentum Diffusion in Magnetosonic versus Alfv\'enic 
Turbulent Fields} 
 
\author{G. \surname{Micha{\l}ek}} 
\author{M. \surname{Ostrowski}} 
 
\institute{Obserwatorium Astronomiczne, Uniwersytet Jagiello\'nski, 
ul.Orla 171, 30-244 Krak\'ow, Poland } 
 
\author{R. \surname{Schlickeiser}} 
 
\institute{Institut f\"ur Theoretische Physik, Lehrstuhl IV: Weltraum und  
Astrophysik, Ruhr-Universit\"at Bochum, D-44780 Bochum, Germany} 
\end{opening}

\begin{ao} 
G. Micha{\l}ek \\ 
Obserwatorium Astronomiczne \\ 
Uniwersytet Jagiello\'nski \\ 
ul.Orla 171 \\ 
30-244 Krak\'ow \\ 
Poland\\ 
E-mail: michalek@oa.uj.edu.pl 
\end{ao} 
 
\begin{document} 
 
\begin{abstract} 
Energetic particle transport in a finite amplitude 
magnetosonic and Alfv\'enic turbulence is considered using Monte 
Carlo particle simulations, 
 which involve an integration of particle 
equations of motion.  
We show that in a low-$\beta$ plasma cosmic 
ray acceleration can be the most important damping process for 
magnetosonic waves. Assuming such conditions we derive the momentum 
diffusion coefficient $D_p$ for relativistic particles in the presence 
of anisotropic finite-amplitude turbulent wave fields, for flat and  
Kolmogorov-type turbulence spectra. We confirm the possibility of 
 larger values of $D_p$ occurring due to transit-time damping resonance 
interaction in the presence of isotropic fast-mode waves in comparison 
to the Alfv\'en waves of the same amplitude (cf. Schlickeiser \& Miller 
1998). The importance of quasi-perpendicular fast-mode waves is stressed 
for the acceleration of high velocity particles. 
\end{abstract} 
 

\section{Introduction} 
 
A number of astronomical objects (extragalactic radio sources, supernova 
remnants, solar flares) emit radiation with non-thermal spectra. These 
emissions are often connected with the existence of a hot turbulent 
magnetized plasma providing conditions for particle acceleration by MHD 
turbulence. It was first shown by Hall \& Sturrock (1967) and by Kulsrud 
\& Ferrari (1971) that charged particles can be accelerated by MHD 
turbulence having wavelengths long compared to the particle gyration 
radius. For example, the importance of a Fermi-like acceleration mechanism 
was considered  
for extragalactic radio sources (Burn 1975; De Young 1976; 
Blandford \& Rees 1978; Achterberg 1979; Eilek 1979). 
Stochastic acceleration by plasma turbulence  
 is often suggested as the acceleration mechanism  
for energetic particles during impulsive solar flares  
 (see, e.g. Miller \& Ramaty 1987; Stein\"acker \& Miller 1992;  
Hamilton \& Petrosian 1992).  
It has been proposed that weak ($\delta B/B_{o} \ll 1$) MHD 
waves can simultaneously energize ions by cascading shear Alfv\'en waves  
(Miller \& Roberts 1995; Miller \& Reames  
1996) as well as electrons by the accompanying cascading fast-mode waves  
through a process known as a transit-time acceleration (Miller et al. 1996). 
The MHD waves are attractive candidates because they can be produced  
by either large-scale restructuring of the magnetic field, which presumably 
occurs during the flare energy release phase, or by a shear in the plasma  
bulk velocity (Roberts et al. 1992), which is likely to be found in regions  
of reconnection-driven plasma outflows (see, e.g., LaRosa \& Moore 1993,  
Forbes 1996). 
 
The interaction between the waves and particles is determined primarily 
by delta functions, $\delta (\omega - k_\parallel v_\parallel - n 
\Omega )$, which select those waves from a spectrum which are 
resonant with a given particle\footnote{ The notation is explained in 
Appendix A}. At such interactions energies vary in a diffusive way, and 
over a long time\-scale a net energy gain results in the process of 
stochastic acceleration. For $n \ne 0$, a particle resonates with those 
waves that are seen at an integral multiple of a gyrofrequency and for 
$n = 0$ at zero frequency. Then, a particle feels a net force which is 
not averaged out by the phase mixing (Stix 1962). The case of $n = 0$ 
provides magnetic analogy to Landau damping and is associated with 
a first-order change in $\delta |\vec{B}|$. This type of coupling is 
not observed in the case of Alfv\'en waves but becomes important for the 
magnetosonic waves containing compressive components of $\delta 
\vec{B}$. The interaction between the particle magnetic momentum and the 
parallel gradient of the magnetic field is called transit-time 
damping (cf. discussions by Lee \& V\"olk 1975, Achterberg 1981, Miller 
et al. 1996). 
 
Recently, Schlickeiser \& Miller (1998) presented a quasi-linear 
derivation of cosmic ray transport coefficients in the presence of MHD 
waves, including isotropic fast-mode turbulence. For the isotropic 
Kolmogorov turbulence they demonstrated that the Fokker-Planck 
coefficients depend both on the transit-time damping and the 
gyro-resonance interactions. For cosmic ray particles with $v \gg V_A$ 
and for vanishing turbulence cross helicity the momentum diffusion 
coefficient in the fast-mode turbulence is mainly determined by the 
transit-time damping contribution, leading to a more efficient  
stochastic acceleration in comparison to the process in the presence of 
pure Alfv\'enic turbulence. 
 
The aim of the present paper is to study the momentum diffusion 
coefficient $D_p$ in the presence of finite amplitude ($ \equiv \delta B \sim 
1$, here we denote $\delta B / B_0 \equiv \delta B$ )  
magnetosonic and Alfv\'en waves. The influence of wave 
anisotropy and of the wave spectrum slope on $D_p$ is considered. In the 
next section we discuss the conditions essential for  fast-mode wave 
damping in the plasma. We show that the low-$\beta$ plasma can provide 
conditions with the main damping process being the cosmic ray particles' 
acceleration. Then, in Section~3, we summarize the results of 
quasi-linear analytic derivations of the momentum diffusion coefficient 
to provide a reference for our numerical modelling. The performed Monte 
Carlo simulations involving derivations of particles trajectories in the 
space filled with finite amplitude fast-mode and Alfv\'en waves are 
described in Section~4. Anisotropic wave distributions are modeled by 
choosing their wave vectors from cones directed along the mean magnetic 
field $\vec{B}_o$. For our simulations we adopt fast-mode or Alfv\'en 
mode turbulence with the flat- ($q = 1$) or the Kolmogorov ($q = 5/3$) 
spectrum within the finite wave vector range ($k_{min}$, $k_{max}$). The 
results are presented and discussed in Section~5. We confirm a 
substantial increase of $D_p$ for the fast-mode waves in comparison to 
the Alfv\'en waves of the same amplitude if the nearly perpendicular 
($\vec{k} \perp \vec{B}_o$) waves are present.

\section{Dissipation of fast-mode waves in a low $\beta $ plasma} 
 
Fast mode waves can effectively accelerate \kr ~particles, if other 
dissipation processes are negligible small. We will demonstrate here 
that in gases with low plasma-beta, $\beta = 2c_s^2/V_A^2 << 1$, this 
requirement is fulfilled. 
 
The equilibrium intensity of plasma waves results from the 
competition of wave generation and wave damping processes. The kinetic 
equation for a 3-dimensional spectral density $\bar{W}_i (\kkv )$, which 
denotes the wave energy density per unit volume of wavenumber space of 
the wave mode $i$, is given by the usual conservation equation 
 
$${{\partial \bar{W}_i(\kkv )} \over \partial t} = -{\partial \over 
\partial \kkv} \cdot \vec{F}(\kkv )\quad, \eqno (2.1)$$ 
 
\noindent 
where the flux 
 
$$\vec{F}(\kkv )=-D{\partial \bar{W}_i(\kkv )\over \partial \kkv} \quad 
\eqno (2.2)$$ 
 
\noindent 
is expressed as a diffusive term with the diffusion coefficient in 
wavenumber space 
 
$$D=k^2/\tau _s(k) \quad \eqno (2.3)$$ 
 
\noindent 
and the spectral energy transfer time scale $\tau _s(k)$. This approach, 
to describe the evolution of turbulence by a diffusion of energy in 
wavenumber space, was pioneered by Leith (1967) in hydrodynamics, and 
subsequently introduced to magnetohydrodynamics by Zhou \& Matthaeus 
(1990). It provides a simple framework to take into account - at least 
approximately - turbulence evolution in space physics applications. Zhou 
\& Matthaeus (1990) present a general transport equation for the wave 
spectral density in the case of isotropic turbulence, which includes terms 
for spatial convection and propagation, nonlinear transfer of energy 
across the wavenumber spectrum, and a source and sink of wave energy. 
 
For isotropic turbulence one obtains for the associated 
one-dimensional spectral density $W_i(k) = 4\pi k^2\bar{W}_i(\kkv )$ the 
simplified diffusion equation 
 
$${\partial W_i\over \partial t} = {\partial \over \partial k} \left [ 
{k^4\over \tau _s(k)} {\partial \over \partial k} \left ( k^{-2}W_i 
\right ) \right] - \; \Gamma _iW_i+\; S_i(k)\quad, \eqno (2.4)$$ 
 
\noindent 
where we included terms for the damping or growth of waves ($\Gamma _i$) 
and a wave energy injection and/or sink term $S_i(k)$. 
 
The spectral energy transfer time scale (or the wavenumber diffusion 
coefficient (2.3)) depends upon the cascade phenomenology. In the 
Kolmogorov treatment the spectral energy transfer time at a particular 
wavelength $\lambda $ is the eddy turnover time $\lambda /\delta v$, 
where $\delta v$ is the velocity fluctuation due to the wave. In the
so-called  
Kraichnan treatment the transfer time is longer by a factor $V_A/\delta 
v$. Both phenomenologies are further discussed in Zhou \& Matthaeus 
(1990) and yield 
 
$$\tau _s(k)\simeq {1\over V_Ak^{3/2}} \cases {\sqrt{{2U_B\over W_i}}& 
(Kolmogorov) \cr {2U_B\over k^{1/2}W_i}& (Kraichnan)\cr }\quad , \eqno 
(2.5)$$ 
 
\noindent 
where $U_B = B_0^2/8\pi $ denotes the energy density of the ordered 
magnetic field. Substituting these transfer time scales into Eq. (2.4), 
and assuming a steady state with no damping, we obtain $W_i = 
W_0k^{-q}$, where $q = 5/3$ for the Kolmogorov case and $q = 3/2$ for 
the Kraichnan phenomenology. The diffusion equation (2.4) in either case 
is nonlinear. 
 
Besides the generation of turbulence by kinetic cosmic ray streaming 
instabilities (e.g. Tademaru 1969),  wave cascading from low 
to high wavenumbers is very often  
an important way of producing broadband wave 
spectra. One possibility, discussed e.g. in the context of solar flares 
(Miller \& Roberts 1995), is that long-wavelength turbulence results 
from the rearrangement of large-scale magnetic fields and/or a shear 
flow instability, so that it is reasonable to assume the deposition of 
wave energy peaked at long wavelength, probably comparable to the 
physical size of the system, as the primary energy release. Cascading as 
described by the nonlinear diffusion term in Eq. (2.4) will then 
transfer this spectral energy to higher wavenumbers, where the waves 
will be able to resonate with progressively lower energy \kr ~particles, 
until they eventually interact with the charged particles in the tail of 
the background thermal distribution. 
 
According to Ginzburg (1970) the damping rate of fast magnetosonic waves 
propagating at an angle $\theta $ with respect to the ordered magnetic 
field in a thermal electron-proton background plasma is given by 
 
$$\Gamma _t = \sqrt{{\pi \beta \over 16}}\, V_A|k| {\sin^2\theta \over 
\cos \theta }F(\beta ,\theta )\quad, \eqno (2.6)$$ 
 
\noindent 
where 
 
$$F(\beta ,\theta )=\sqrt{{m_e \over m_p}}+\; 5\exp [-(\beta \cos 
^2\theta )^{-1}] \eqno (2.7)$$ 
 
\noindent 
depends on the plasma beta of the background plasma 
 
$$\beta ={2c_S^2\over V_A^2}={8\pi n_em_pk_BT_e\over B_0^2}\quad. \eqno 
(2.8)$$ 
 
\noindent 
The function $F(\beta ,\theta )$ exhibits an almost perfect "flip-flop" 
behaviour with constant values $F(\beta \le \beta _c)\simeq 
\sqrt{m_e/m_p} \simeq 1/43$ and $F(\beta >\beta _c)\simeq 5$ below and above 
$\beta _c = 0.5\cos^{-2}\theta \ge 0.5$. 
 
The other relevant wave damping process is the transit-time damping 
acceleration of isotropically distributed cosmic rays. According to 
Achterberg (1981) the corresponding damping rate is given by 
 
$$\Gamma _r = {\pi V_A\over 4c}\, V_A|k| {\sin^2\theta \over \cos 
\theta } \left [ 1-{V_A^2\over c^2\cos^2\theta }\right ]^2\, {U_p\over 
U_B}\quad , \eqno (2.9)$$ 
 
\noindent 
where $U_B$ and $U_p$ denote the energy densities of the background 
magnetic field and the cosmic ray particles, respectively. 
 
For small wave intensities we can neglect the cascading of waves, so 
that according to Eq. (2.4) the equilibrium fast-mode intensity is 
simply given by 
 
$$W_f(k) = {S_f(k)\over \Gamma _t+\Gamma _r}\quad . \eqno (2.10)$$ 
 
\noindent 
Calculating following Achterberg (1981) the ratio of the two damping rates, 
 
$$R\equiv {\Gamma _t\over \Gamma _r} = {U_B \over U_p} \left [ 
1-{V_A^2\over c^2\cos^2\theta } \right ]^{-2}\, {c\over \pi 
^{1/2}V_A}\, \beta^{1/2}F(\beta ,\theta )\quad , \eqno (2.11)$$ 
 
\noindent 
we find that this ratio is independent of wavenumber $k$, and for small 
plasma beta $\beta < 0.5$ almost independent from the propagation angle 
$\theta $ since $V_A << c$. The ratio is well approximated by 
 
$$R(\beta <0.5)\simeq {U_B\over U_p} {c\over V_A}\, \sqrt{ {\beta \over 
1836 \pi } } = {U_{B} \over U_p} {c\over c_S}\, {\beta \over \sqrt{ 
3672 \pi } } \quad . \eqno (2.12)$$ 
 
\noindent 
Values of $R < 1$ indicate that the fast-mode wave energy is damped by 
accelerating cosmic ray particles. The condition $R < 1$ translates into 
a condition for the plasma beta 
 
$$\beta \le 0.01 \sqrt{T_5}{ U_p \over U_B}\quad ,\eqno (2.13)$$ 
 
\noindent 
where $T_5 = (T/10^5$K) denotes the plasma temperature in units of 
10$^5$K. In the case of equipartition between magnetic field and cosmic 
rays, $U_p/U_B \simeq 1$ we find values of $R$ less unity if $\beta \le 
0.01 \sqrt{T_5}$. In conclusion, in a low $\beta $-plasma the fast-mode 
waves are predominantly dissipated by accelerating cosmic rays by 
transit-time damping. The inclusion of wave cascading does not 
qualitatively modify this conclusion as the numerical solutions of 
Miller et al. (1996) indicate.

\section{Quasi-linear momentum diffusion coefficient} 
 
The quasi-linear theory treats the effect of the weakly perturbed 
magnetic field as perturbations of orbits of particles moving in the 
average background field. Schlickeiser (1989) considers quasilinear 
transport and acceleration parameters for cosmic ray particles 
interacting resonantly with the Alfv\'en waves propagating along  
the average magnetic field. The transport equation can be derived from 
the Fokker-Planck equation by a well-known approximate scheme (Jokipii 
1966, Hasselmann \& Wibberenz 1968) which is commonly referred to as the 
diffusion-convection equation for the pitch-angle averaged phase space 
density.  The electromagnetic fields generated by MHD 
waves enter into the equation through the Lorentz force term. 
For fast ($v \gg V_A$) 
cosmic ray particles, a vanishing cross helicity state of the Alfv\'en 
waves and the power-law turbulence spectrum with $q \geq 1$, 
 
$$ \bar{W}_i(\kkv )=  
\bar{W}_o^i(\delta 
B_{i})^2 k^{-q}~~~for~~~(k_{min} < k < k_{max}) \quad \eqno(3.1)$$ 
 
\noindent 
the momentum diffusion coefficient  
 
$$D_p={\pi \bar{W}_o\over q(q+2)} \Bigl{(}{\delta B \over B } 
\Bigr{)}^2  |\Omega|^{2-q} {V_A^2  p^2  \over v^{3-q}} \quad , 
\eqno(3.2)$$ 
 
\noindent 
where 
 
$$\bar{W}_o={1-q \over k_{max}^{1-q}-k_{min}^{1-q} } ~~~{\rm for}~~~ 
q>1$$ 
 
\noindent 
and 
 
$$ \bar{W}_o=\left ( \ln {k_{max} \over k_{min}} \right )^{-1} ~~~{\rm 
for}~~~q=1\quad .$$ 
 
Recently, Schlickeiser \& Miller (1998) considered cosmic ray particles 
interacting with oblique fast-mode waves propagating in a low-$\beta$ 
plasma. In the cold plasma limit the fast and slow magnetosonic waves 
degenerate to the fast-mode waves with the dispersion relation $\omega^2 
= V_A^2 k^2 $. The momentum diffusion coefficient (in the original paper 
`$a_2$' for our $D_p$) and the spatial diffusion coefficient 
$\kappa_\parallel$ can be calculated as respective pitch angle averages 
of the Fokker-Planck coefficient $D_{\mu \mu}(\mu,p)$. Adopting 
isotropic fast-mode turbulence with a power-law turbulence spectrum 
(3.1) they obtained 
 
$$D_{\mu \mu} = {\pi | \Omega | \bar{W}_o(1-\mu^2 ) \over 4} 
\Bigl{(}{\delta B \over B } \Bigr{)}^2 
(R_L)^{q-1}[f_T(\mu)+f_G(\mu)] \quad , \eqno(3.3)$$ 
 
\noindent 
where $D_{\mu \mu}$ includes a sum of the transit-time damping ($f_T$) 
and the gyroresonance ($f_G$) interaction contributions. A considered 
form of 
$f_T$ admits the pitch angle scattering by transit-time damping of 
super-Alfv\'enic particles with pitch-angles contained in the range 
$\epsilon \leq | \mu | \leq 1$, where $\epsilon \equiv V_A/v$. In the 
interval $| \mu | \le \epsilon$, where no transit-time damping occurs, 
the gyroresonance interactions provide a small but finite contribution 
to the particle scattering rate. As a result the momentum diffusion 
coefficient $D_p$ is mainly determined by the transit-time damping 
interactions and the spatial diffusion coefficient by the gyroresonance 
interactions. For $1 \leq q \le 2$ fast-mode turbulence spectra (Eq. 
3.1) they obtained 
 
$$D_p \simeq {\pi \bar{W}_o C_1 \over 4} \Bigl( {\delta B \over B } 
\Bigr)^2 | \Omega | (R_L)^{q-1} {V_A^2 p^2 \over v^2  }\, \ln {v \over 
V_A} \quad , \eqno(3.4)$$ 
 
\noindent 
where 
 
$$C_1 = 2^{1-q} { q \Gamma (q) \Gamma (2-{q \over 2}) \over (4 - q^2) 
\Gamma^3 (1+ {q \over 2}) } \quad . \eqno(3.5)$$ 
 
\section{Description of simulations} 
 
The approach applied in the present paper for modelling the particle 
momentum diffusion is based on numerical Monte Carlo simulations. The 
general procedure is quite simple: test particles are injected at random 
positions into a turbulent magnetized plasma and their trajectories are 
followed by integration of particle equations of motion. Due to the 
presence of waves, particles move diffusively in configuration and momentum 
space . By 
averaging over a large number of trajectories one derives the diffusion 
coefficients for turbulent wave fields. In the simulations we consider 
relativistic particles with $v \gg V_A$ and use dimensionless units 
(cf. Appendix A): $\delta B \equiv \delta B / B_o$ for magnetic field 
perturbations, $1 / \Omega_o$ for time, $k/k_{res}$ for wave vectors and 
$p_o^2 \Omega_o$ for the momentum diffusion coefficient. 
Below, in all our simulations we adopt $V_A = 10^{-3}c$ and we consider  
mono-energetic particles with velocity $v = 0.99c$. This particular  
choice allows us to compare the effects of different types of turbulence 
on the particle momentum diffusion coefficient and, due to the large $v$, 
to evaluate the role of wave anisotropy in the transit-time damping  
interactions.

\subsection{The Wave Field Models}  
  
In the modelling we consider a superposition of 384 MHD waves 
propagating oblique to the average magnetic field $\vec{B}_o \equiv B_o 
\hat{\bf e}_z$. The wave propagation angle with respect to $\vec{B}_o$ 
is randomly chosen from a uniform distribution within a cone 
(`wave-cone') along the mean field. For a given simulation two symmetric 
cones are considered centered along $\vec{B}_o$, with the opening angle 
$2\alpha$, directed parallel and anti-parallel to the mean field 
direction. The same number of waves is selected from each cone in order 
to model the symmetric wave field. Related to the  i-th wave, 
the magnetic 
field fluctuation vector $\delta \vec{B}^{(i)}$ is given in the form: 
  
$${\delta \vec{B}^{(i)} = \delta \vec{B}_o^{(i)} \sin (\vec{k}^{(i)}  
\cdot \vec{r} - \omega^{(i)} t ) } \quad , \eqno(4.1)$$ 
 
\noindent 
where $\delta \vec{B}_o^{(i)}$ is a constant vector selected for a 
given wave. The electric field fluctuation related to a particular wave 
is given as $\delta \vec{E}^{(i)} = - \vec{V}^{(i)} \times \delta 
\vec{B}^{(i)}$. For Alfv\'en waves (`A') we use the dispersion relation 
  
$$ \omega_A^2 = k_\|^2 V_A^2 \quad , \eqno(4.2)$$  
  
\noindent 
where $V_A = B_o / \sqrt{4\pi \rho}$ is the Alfv\'en velocity in the 
field $B_o$. The wave magnetic field polarization is defined by the 
formula 
  
$$ \delta \vec{B}_A = \delta B_A (\vec{k}, \omega_A) \, (\vec{k} \times 
\hat{e} _z) \, k_\perp^{-1} \qquad . \eqno(4.3)$$ 
  
\noindent  
In a low-$\beta$ plasma the fast-mode magnetosonic waves (`M') propagate 
with the Alfv\'en velocity and the respective relations are: 
  
$$ \omega_M^2 = k^2 V_A^2 \qquad , \eqno(4.4)$$ 
  
$$\delta \vec{B}_M = \delta B_M ( \vec{k} ,\omega_M) \, (\vec{k} \times 
(\vec{k} \times \hat{e}_z)) \, k^{-1} \, k_{\perp}^{-1} \qquad . 
\eqno(4.5)$$ 
 
\noindent 
One 
should be aware of the fact that the considered turbulence model is 
unrealistic at large $\delta B$ and the present results can not be 
considered as the exact quantitative ones. In particular, in the 
presence of a finite amplitude turbulence the magnetic field pressure is 
larger than the mean field pressure and the wave phase velocities can be 
greater than $V_A(B_o)$ assumed here. 
 
\subsection{Spectrum of the turbulence} 
 
In our simulations we consider the power-law turbulence spectrum in the 
wave vector range ($k_{min}$, $k_{max}$). The amplitude of the 
irregular component of the magnetic field, obtained from the energy 
density defined in equation (3.1), can be written as 
  
$$ \delta B(k)=\delta B(k_{min}) \Big( {k \over k_{min}} \Big)^{- q /2} 
\quad , \eqno(4.6)$$ 
 
\noindent 
where $k_{min} = 0.08$ ($k_{max} = 8.0)$ corresponds to the considered 
longest (shortest) wavelength, respectively, and $q$ is the wave 
spectral index. In the present simulations we consider the flat spectrum 
with $q = 1$ and the Kolmogorov spectrum with $q = 5/3$. We included the 
flat spectrum because of our earlier simulations (Michalek \& Ostrowski 
1996), where the Alfv\'en waves with $q = 1$ were considered. It provides  
also a convenient limiting reference for comparison of results for different 
wave spectra flatter than the Kolmogorov one (e.g. the Kraichnan spectrum  
or the perturbations' spectra at wavelengths longer than the ones  
characteristic for the inertial range of the MHD cascade). On the 
other hand such a turbulence spectrum is very convenient for numerical 
simulations due to presence of a substantial power in short waves. For 
the flat spectrum the wave vectors are drawn in a random way from the 
respective ranges: $2.0 \le k \le 8.0$ for `short' waves, $0.4 \le k \le 
2.0$ for `medium' waves and $0.08 \le k \le 0.4$ for `long' waves. The 
respective wave amplitudes $\delta B_o$ are drawn in a random manner so 
as to keep constant 
 
$$ \sum_{i=1}^{384}(\delta B_o^{(i)})^2 \equiv \delta B^2 , \qquad 
\eqno(4.7)$$ 
 
\noindent 
where $\delta B$ is a model parameter, and, separately in all mentioned 
wave-vector ranges 
 
$$ \sum_{i=1}^{128}(\delta B_o^{(i)})^2 \equiv 
\sum_{i=129}^{256}(\delta B_o^{(i)})^2 \equiv \sum_{i=257}^{384}(\delta 
B_o^{(i)})^2 \equiv {\delta B^2  \over 3}\quad. \eqno(4.8)$$ 
 
Thus the wave energy is uniformly distributed over the considered 
wave-vector ranges. As a second, more realistic turbulence model we 
consider  one involving the Kolmogorov spectrum. The observed 
spectra in  interplanetary space often have such a form (e.g. Jokipii 
1971) and this case is most often discussed in the literature. Here all 
384 wave vectors are drawn in a random manner from the whole considered 
range ($0.08 \le k \le 8.0$), but the amplitudes $\delta B^{(i)}$ are 
fitted according to the Kolmogorov distribution (Eq. 4.6 with $q = 5/3$) 
and scaled to keep the formula (4.7) valid. In such turbulence most of 
energy is carried by long waves. 
 
In the discussion below we will consider four different turbulence 
fields labeled as follows:\\ 
i.~~~Alfv\'en waves with the flat spectrum - AF,  \\ 
ii.~~Alfv\'en waves with the Kolmogorov spectrum - AK, \\ 
iii.~Fast-mode waves with the flat spectrum - MF, \\ 
iv.~Fast-mode waves with the Kolmogorov spectrum - MK. \\

\section{Results} 
 
In the present numerical simulations we consider a number of physical 
situations -- turbulence models with finite wave amplitudes and 
a changing degree of wave anisotropy -- not described by analytic means. 
The results are illustrated in figures 1 and 2. First,  
 the simulated momentum diffusion coefficients, $D_p$, 
for the Alfv\'en and the magnetosonic turbulence are presented in Fig.~1.:  
in the 
upper panels variation of $D_p$ versus the perturbation amplitude 
$\delta B$ for the Alfv\'en turbulence and in the middle panels for the 
fast-mode turbulence. To enable comparison these results are 
superimposed in bottom panels. The derived momentum diffusion 
coefficients are given for different wave-cone opening angles. The 
models with $\alpha = 0^\circ$, $\alpha = 40^\circ$ and $\alpha = 
90^\circ$ represent degenerated one-dimensional, anisotropic and 
isotropic wave vector distributions, respectively. For  clarity, only  
the results for the mentioned three values are presented, but  we also 
performed simulations  for other intermediate wave cone openings: 
$\alpha = 30^\circ$ and $60^\circ$. The results of these simulations 
are consistent with the ones described below. 
 
At Fig.~1, in all cases a systematic increase of $D_p$ with $\delta B$ 
can be observed. Usually, it follows the quasi-linear relation $D_p \propto 
\delta B^2 $, except for the magnetosonic waves with $\alpha = 90^\circ$, 
where the increase rate seems to be smaller at least at large 
amplitudes, where it is roughly $\propto \delta B^{1.5}$. 
 
\begin{figure*} 
\vspace{17cm} 
\includegraphics{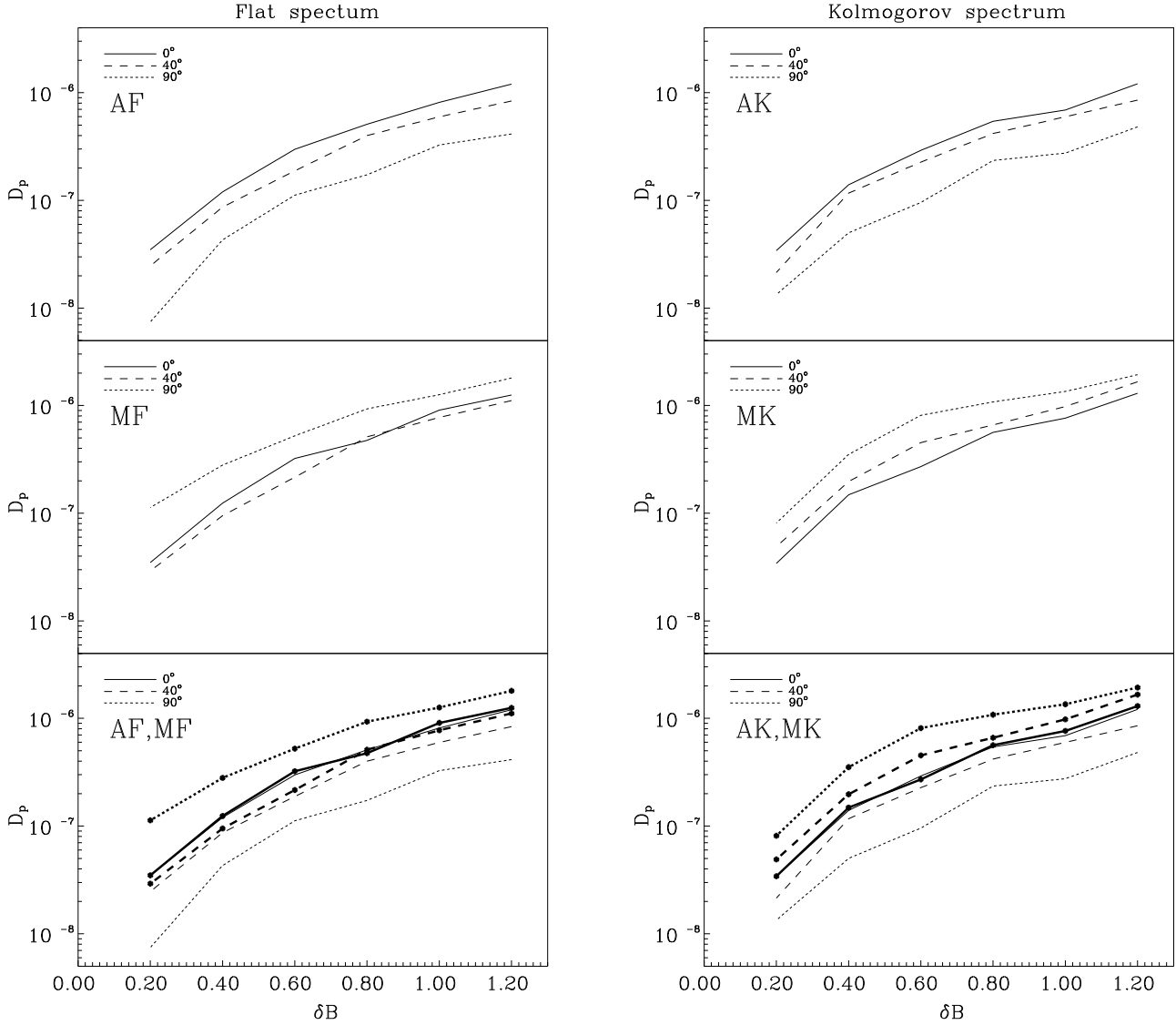} 
\caption[]{Variation of the momentum diffusion coefficient $D_p$, given 
in units of $p_{o}^{2}\Omega_{o}$, versus 
the perturbation amplitude $\delta B$ and the wave propagation 
anisotropy. The results are presented for the flat and the Kolmogorov 
spectra, separately for the Alfv\'en wave and the fast-mode turbulence. 
For comparison the results for the Alfv\'en wave turbulence (thin lines) 
and the fast-mode turbulence (thick lines with indicated simulation 
points) are superimposed at the bottom panels.} 
\end{figure*} 
 
For the Alfv\'en wave turbulence an increase of the wave-cone opening 
angle $\alpha$ providing waves with smaller phase velocities leads to 
decreasing $D_p$. The trend is independent of the considered turbulence 
amplitude and the considered spectrum. In the case of fast-mode waves a 
more complicated relation is observed. For the flat turbulence spectrum 
a small increase of the angle $\alpha$ does not lead to a significant 
variation of $D_p$, and the change is to smaller values of $D_p$, like 
for the Alfv\'en waves. Only the appearance of waves with wave vectors 
nearly perpendicular to the mean magnetic field changes this trend 
leading to an increase of $D_p$ (by a factor of three 
for $\alpha = 90^\circ$ in comparison to $\alpha = 0^\circ$ 
at $\delta B = 0.2$). For the Kolmogorov turbulence the wave cone 
opening-angle increase in the MK model is always followed by an increase 
of $D_p$ (in our simulations small values of $\alpha$, between $0^\circ$ 
and $30^\circ$, were not considered) and again it reaches the maximum in 
the presence of perpendicular waves. 
 
In order to explain this non-monotonic behavior one can refer to the 
quasi-linear derivations of Schlickeiser \& Miller (1998). For $n = 0$ 
the resonance condition for the transit-time damping may be written as 
$v_\| = \omega / k_\|$ and, with the dispersion relation (4.4), $V_A / 
v_\| = k_\| / k$. It is clear that for $V_A \ll v$ particles can 
effciently interact with waves at a wide range of $v_\|$ only if the 
waves with $ k_{\perp} \gg k_\|$ are present. This fact can explain the 
effective transit-time damping interactions for magnetosonic waves with 
isotropic distribution. Then, the observed $D_p$ increase results 
mainly from the presence of a small fraction of waves propagating 
quasi-perpendicular to the mean magnetic field. One should note that our 
simulations were performed for relativistic particles with $v=0.99c \gg 
V_A$, where the effect is quite pronounced. 
 
In order to verify this behaviour in more detail we considered 
interaction of such relativistic particles with waves propagating in 
narrow pitch angle ranges with respect to the mean magnetic field: we 
performed simulations for wave vector inclinations selected from the 
angular ranges ($\pm 5^\circ$) around the cones with $\alpha=0^\circ$, 
$45^\circ$ and $85^\circ$ (Fig.~2). It is clear from the figure that 
only the presence of waves with wave vectors perpendicular to the mean 
magnetic field leads to a significant increase of $D_p$. For these waves 
the transit-time damping resonance provides the wave-particle coupling, 
while the waves propagating nearly parallel to the average magnetic 
field contribute preferentially due to the gyroresonance. Of course, our 
reference to the calculations involving various resonances is 
only approximate, as in the presence of high amplitude waves the 
quasi-linear delta form of interaction changes into the `broadened 
resonance', as discussed by Karimabadi et al. (1992). The simulation 
errors can be evaluated from comparison at Fig-s~1,2 of the `M' and `A' 
results for $\alpha = 0^\circ$, which should coincide. 
 
\begin{figure} 
\vspace{11cm} 
\includegraphics{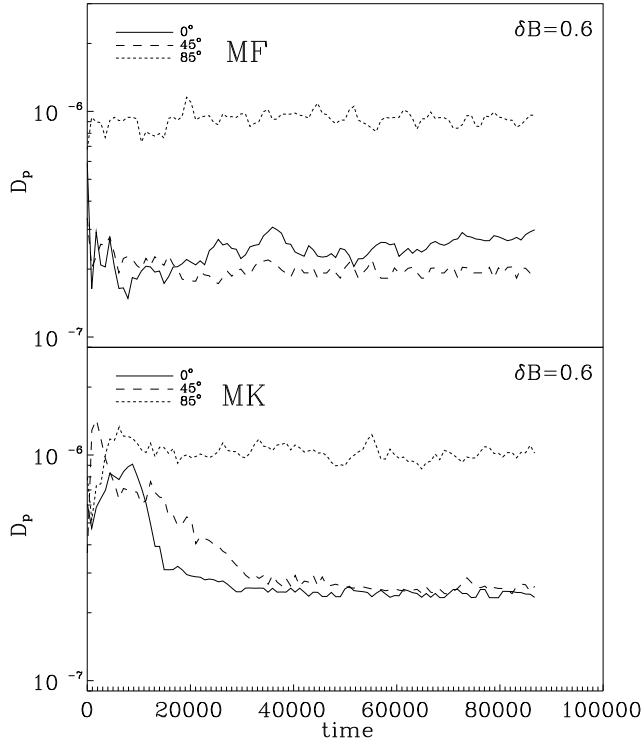} 
\caption[]{Examples of the simulated $D_p$, 
given in units $p_{o}^{2}\Omega_{o}$, 
versus the simulation time 
for waves selected from the respective ($0^\circ\pm 5^\circ$, 
$45^\circ\pm 5^\circ$, $85^\circ \pm 5^\circ$) pitch angle ranges for 
both types of turbulence spectrum at $\delta B = 0.6$.} \end{figure} 
\noindent 
 
\section{Final remarks} 
 
We considered the momentum diffusion coefficient $D_p$ in the presence of 
oblique Alfv\'en and magnetosonic waves with a wide range of amplitudes, from 
medium, up to non-linear ones. The influence of the degree of wave 
anisotropy and the waves' spectral index was also studied. 
As expected, in all cases a systematic increase of $D_p$ with the wave 
amplitude is observed. We confirm a substantial increase of $D_p$ for 
the fast-mode waves in comparison to the Alfv\'en waves of the same 
amplitude, if nearly perpendicular fast-mode waves are present. The 
effect is caused by the transit-time damping interactions which occur in 
the presence of magnetosonic waves containing a compressive component of 
$\delta \vec{B}$. For the isotropic fast-mode Kolmogorov 
turbulence (cf. Eq-s 3.2,4) $D_p$ is expected to be on a factor of  
$ln(V/V_A)$, multiplied by a term depending on the index $q$,  
larger than the one for  Alfv\'en waves propagating parallel to 
$\vec{B}_o$. The ratio of these values  
is about three in our simulations for small $\delta B$, which 
confirms (within the simulation errors) the quasi-linear result. In 
simulations for the isotropic Kolmogorov spectra $D_p^{MK}/D_p^{AK} 
\approx 10$ is roughly constant for a wide range of waves amplitudes. 
For the isotropic MK and MF models one observes that the transit-time 
damping scattering weakens with an increasing $\delta B$. For these 
models,  $D_p \propto (\delta B)^{1.5}$, when some particles can 
become timely trapped between the compressive waves and the acceleration 
due to the transit-time damping becomes less efficient. 
 
Fast-mode waves can effectively accelerate cosmic ray particles if other 
dissipation processes heating the plasma  
are negligibly small. In Section~2 we 
demonstrated that for a low beta-plasma, $\beta <<1$, the fast-mode 
waves can be predominantly dissipated by accelerating cosmic rays 
through the transit-time damping mechanism. In addition, the 
momentum diffusion enhancement due to presence of fast-mode waves could 
also work effectively in a volume with  the turbulence generation 
force acting. For example, in the  
vicinity of a strong shock, where the conditions 
with $U_p \gg U_B$ can occur (cf. Drury 1983), or in a region of 
magnetic field reconnection the required fast-mode perpendicular waves 
are expected to be effectively created. One should remember that the 
waves' phase velocities are expected to be larger in realistic 
turbulence than $V_A(B_o)$ assumed here. Therefore our results for 
large $\delta B$ should rather  be considered  as the lower limits for 
$D_p$. 
 
\begin{acknowledgements} 
MO thanks Prof. Richard Wielebinski for a kind invitation to 
the Max-Planck-Institut f\"ur Radioastronomie in Bonn, where part of the 
present work was done. We are grateful to Dr. Hui Li for useful 
discussions and to Dr. Horst Fichtner for correcting the final version 
of the paper.  GM \& MO acknowledge the Komitet Bada\'n Naukowych support 
through the grant PB 179/P03/96/11. Simulations were performed in ACK  
CYFRONET KRAK\'OW (KBN/SPP/UJ/006/1997). 
\end{acknowledgements} 
 
\section*{References} 
 
\parskip=0pt 
\parindent=1cm 
 
Achterberg A., 1979, A{\&}A, 76, 276 \\ 
Achterberg A., 1981, A{\&}A, 97, 259\\ 
Blandford R.D., Rees M.I., 1978, Physica Scripta, 17, 3543\\ 
Burn B.I., 1975, A{\&}A, 45, 435\\ 
De Young D.S., 1976, ARAA, 14, 447\\ 
Drury L.O'C., 1983, Rep. Prog. Phys., 46, 973\\ 
Eilek J.A., 1979, ApJ, 230, 373\\ 
Forbes T.G., 1996, High-Energy Solar Physics, ed. R.Ramaty, \par 
N.Mandzhavidze \& X.-M.Hua (New York: AIP), 275\\ 
Ginzburg, V. L., 1970, {\it The propagation of electromagnetic \par  
waves in plasmas}, Pergamon Press, Oxford\\ 
Hall D.E., Sturrock P.A., 1967, Phys. Fluids, 10, 2620\\ 
Hamilton R.J., Petrosian V., 1992, ApJ, 398, 350\\ 
Hasselmann K., Wibberenz G., 1967, Z. Geophys., 34, 353\\  
Jokipii J.R., 1966, ApJ, 146, 480\\ 
Jokipii J.R., 1971, Rev. Geophys. Space Phys., 9, 27\\ 
Karimabadi H., Krauss-Varban D., Terasawa T., 1992, JGR\par 
97, 13853\\ 
Kulsrud R.M., Ferrari A., 1971, ASS, 12, 302\\ 
LaRosa T.N., Moore R.L., 1993, ApJ, 418, 912\\ 
Lee M.A., V\"olk H.J., 1975, ApJ, 198, 485\\ 
Leith, C. E., 1967, Phys. Fluids, 10, 1409\\ 
Melrose D.B., 1974, Sol. Phys., 37, 353\\ 
Micha\l ek G., Ostrowski M., 1996, Nonlinear Processes \par 
in Geophysics 3, 66 \\ 
Micha{\l}ek G., Ostrowski M., 1997, A{\&}A, 326,793 \\  
Miller J.A., Ramaty R., 1987, Sol. Phys., 113, 195 \\ 
Miller J.A., Roberts, D.A., 1995, ApJ, 452, 912\\ 
Miller J.A., LaRosa T.N., Moore R.L., 1996, ApJ, 461, 445\\ 
Miller J.A., Reames D.V. 1996,  High-Energy Solar Physics, \par ed. R.Ramaty, 
N.Mandzhavidze \& X.-M.Hua (New York: AIP), 450\\ 
Ramaty R., Kozlovsky B., Lingenfelter R.E., 1979, ApJS, 40, 487\\ 
Roberts D.A., Goldstein M.L., Mattheaus W.H., Ghosh S., 1992 \par 
JGR, 97, 17115\\ 
Schlickeiser R., 1989, ApJ, 336, 243\\ 
Schlickeiser R., Miller J.A., 1998, ApJ, 492, 352 \\  
Stein\"acker J., Miller J.A., 1992, ApJ, 393, 764\\ 
Stix T.H., 1962, {\it The Theory of Plasma Waves} \par 
McGraw-Hill, New York\\ 
Tademaru E., 1969, ApJ, 158, 958\\ 
Zhou, Y., Matthaeus, W. H., 1990, JGR, 95, 14881\\ 
\appendix

\section{Summary of notation} 
${\bf B} = {\bf B}_0 + \delta {\bf B}$ -- a magnetic induction vector \\ 
${\bf B}_0$ -- a regular component of the background magnetic field \\ 
$c$ -- the light velocity \\ 
$c_{s}$ -- a velocity of sound \\ 
$D_p \equiv a_{2} $ -- a momentum diffusion coefficient \\ 
$D_{\mu \mu}$-- a pitch-angle diffusion coefficient \\ 
$e$ -- a particle charge \\ 
${\bf E}$ -- an electric field vector \\ 
$\vec{F}(\kkv )$ -- an energy flux of waves \\ 
$k$ -- a wave vector \\ 
$k_{res} = 2 \pi / r_g$ -- a resonance wave vector \\ 
$k_\parallel$ -- a wave vector component along $\vec{B}_o$ \\ 
$m_e$ -- the electron mass \\ 
$m_p$ -- the proton mass \\ 
$m$ -- a particle mass \\ 
$n_e$ -- a concentration of electrons \\ 
${\bf p}$ -- a particle momentum vector $(|\vec{p}|=p)$ \\ 
$q$ -- a wave spectral index \\ 
$r_g \equiv c/|\Omega|$ -- a particle maximum Larmour radius \\ 
$R_L = v / |\Omega|$ -- a particle Larmour radius \\ 
$R \equiv \Gamma_t / \Gamma_r$ \\ 
$S_i (k)$ -- an injection or a sink term for wave energy \\ 
$U_B \equiv B_0^2 / 8 \pi$ -- the energy density of the ordered magnetic 
field \\ 
$U_p$ -- the energy density of cosmic ray particles \\ 
${\bf v} $ -- a particle velocity vector $(|\vec{v}| \equiv v)$ \\ 
$ V_A $ -- the Alfv\'en velocity in the field $B_0 $ \\ 
$v_\parallel$ -- a particle velocity along the mean magnetic field \\ 
$\bar{W}_i(\kkv )$ -- a 3-dimensional spectral density for the wave mode 
`$i$' \\ 
$\alpha$ -- a wave-cone opening angle \\ 
$\beta = 2 c_s^2 / V_A^2$ -- a plasma-beta parameter \\ 
$\delta {\bf B}$ -- a turbulent component of the magnetic field \\ 
$\delta v$ -- a characteristic velocity fluctuation of the wave \\ 
$\gamma \equiv (1-v^2/c^2)^{-1/2} $ -- the particle Lorentz factor \\ 
$\Gamma_i$ -- a damping or a growth term for the wave mode `$i$' \\ 
$\Gamma_r$ -- a damping rate due to the transit-time acceleration \\ 
$\Gamma_t$ -- a damping rate in a thermal electron-proton plasma \\ 
$\omega$ -- a wave frequency \\ 
$\omega_A$ -- a frequency of the Alfv\'en wave \\ 
$\omega_{M}$ -- a frequency of the magnetosonic wave \\ 
$\Omega \equiv e B / \gamma m c$ -- a particle angular velocity \\ 
$\Omega_o \equiv eB_o/mc$\\ 
$\mu \equiv \cos \Theta$ \\ 
$\theta$ -- a wave propagation angle with respect to $\bf{B}_o$ \\ 
$\Theta$ -- a momentum pitch-angle with respect to ${\bf B}_0$ \\ 
$\tau _s(k)$ -- a spectral energy-transfer time scale \\ 
$\epsilon\equiv V_A / v$ \\ 
 
\end{document}